\documentclass[12pt]{article}

\usepackage{lineno}

\usepackage[margin=1in]{geometry}
\usepackage{amssymb,amsmath,latexsym}                        
\usepackage{graphicx}
\usepackage{cite}
\usepackage{xcolor}
\usepackage{float}

\usepackage{amsthm}
\usepackage{multicol}
\usepackage{authblk}
\usepackage[colorlinks=true,linkcolor=blue, citecolor=blue, urlcolor=blue]{hyperref}

\usepackage{adjustbox}
\usepackage{bbm}
\usepackage{mathtools}

\date{\today} 

\begin{document}

\title{The Impact of Incorporating Shell-corrections\\ to Energy Loss in Silicon}
\author[1,2]{Fuyue Wang}
\author[3]{Su Dong}
\author[2]{Benjamin Nachman}
\author[2]{Maurice Garcia-Sciveres}
\author[3]{Qi Zeng}
\affil[1]{\normalsize\it Department of Engineering Physics, Tsinghua University, Key Laboratory of Particle and Radiation Imaging, Ministry of Education, Beijing 100084, China}
\affil[2]{\normalsize\it Physics Division, Lawrence Berkeley National Laboratory, Berkeley, CA 94704, USA}

\affil[3]{\normalsize\it SLAC National Accelerator Laboratory, Stanford University, Menlo Park, CA 94025, USA}

\maketitle


\begin{abstract}
Modern silicon tracking detectors based on hybrid or fully integrated CMOS technology are continuing to push to thinner sensors.  The ionization energy loss fluctuations in very thin silicon sensors significantly deviates from the Landau distribution.  Therefore, we have developed a charge deposition setup that implements the Bichsel straggling function, which accounts for shell-effects.  This enhanced simulation is important for comparing with testbeam or collision data with thin sensors as demonstrated by reproducing more realistically the degraded position resolution compared with na\"{i}ve ionization models based on simple Landau-like fluctuation.  Our implementation of the Bichsel model agrees well with the multipurpose photo absorption ionization (PAI) model in Geant4 and is significantly faster.  The code is made publicly available as part of the Allpix software package in order to facilitate predictions for new detector designs and comparisons with testbeam data.

\end{abstract}

\section{Introduction}
\label{sec:intro}


The innermost layers of most modern collider tracking detectors are silicon pixels, using either hybrid modules or fully integrated CMOS technology.  Requirements on the material budget and radiation hardness are pushing sensors to become thinner.  Energy fluctuations in thick sensors is well-described by the Landau-Vavilov distribution~\cite{Landau:1944if,Vavilov:1957zz}.  However, when the sensor is sufficiently thin so that the number of collisions is small and the deposited energy still has the imprint of the shell structure of the silicon atom, the Landau-Vavilov distribution is not a good approximation.   For thin sensors, the Bichsel straggling function is more complete and has been shown to reproduce measured energy losses~\cite{Bichsel:1988if}.

State-of-the-art simulation of material interactions is provided by the Geant4~\cite{AGOSTINELLI2003250} toolkit for most high energy physics experiments, including testbeam simulations.  The experiments at the LHC\footnote{While not used during particle propagation through the detector, the Bichsel model is used by the CMS experiment in a dedicated standalone simulation for a lookup table of charge sharing~\cite{Swartz:2002kda}.} most commonly use variations of the \texttt{EMstandard}\footnote{The actual energy loss routine is the \texttt{Universal Function} which follows the Urb{\'a}n model~\cite{LassilaPerini:1995np}.  This two-state model uses Rutherford ($\propto 1/E^2$) cross-sections with a fudge-factor to match the most probable dE/dx.  The width of the distribution is inflated for small thicknesses with an ad-hoc correction.} physics process list for Geant4 simulation~\cite{ALLISON2016186}. This list does not include shell electron effects and due to its simplistic model, has a much faster execution time.  The physical processes incorporated in \texttt{EMstandard} are an excellent model for thick sensors, but as they result in Landau-Vavilov-like distributions for the energy loss, they are not applicable for thin sensors.  Geant4 does include a physics list with a more detailed energy loss model: the Photo Absorption Ionization (\texttt{PAI}) model~\cite{Allison:1980vw}. The Geant4 implementation of the PAI model~\cite{Apostolakis:2000yu} is based on a corrected table of photo-absorption cross section coefficients. The simulated energy loss is in good agreement with the experiment data on energy loss for moderately thin sensors.  However, the \texttt{PAI} model is not widely used by the major experiments because it is computationally expensive.  Furthermore, like \texttt{EMstandard}, the \texttt{PAI} model is a generic approach that works for various elements while the Bichsel model has been refined based on extensive specific knowledge about silicon.

\begin{figure}[h!]
	\includegraphics[width=0.95\textwidth]{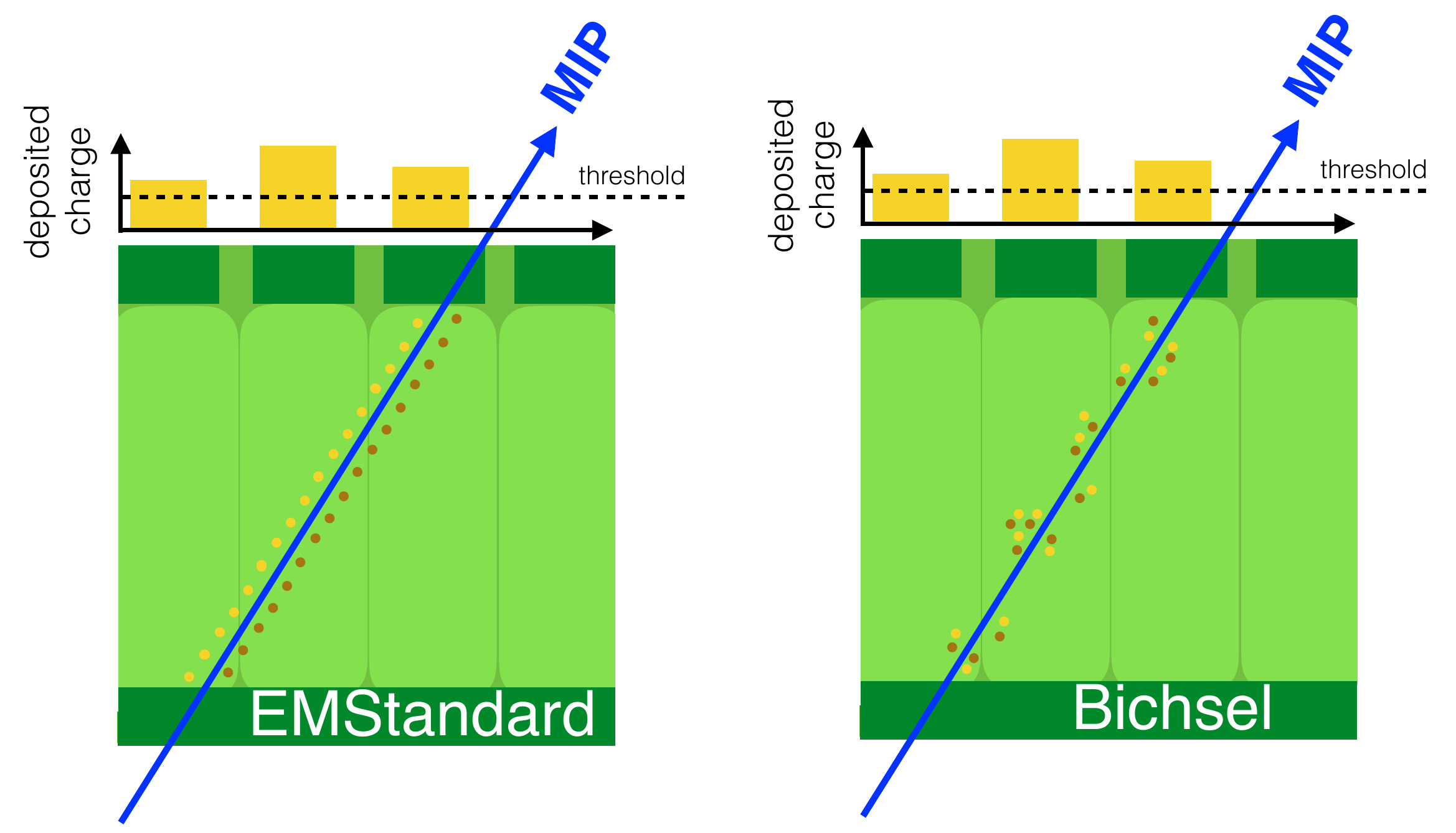}
	\caption{Schematic diagrams of the energy deposition inside a pixel module using the \texttt{EMstandard} (left) and the Bichsel model (right).}
	\label{fig:schematic}
\end{figure}

A dedicated implementation of the Bichsel straggling function has recently been implemented as part of the charge deposition model for the ATLAS detector~\cite{qithesis}.  In this model, the energy fluctuations significantly deviate from those introduced by \texttt{EMstandard}.  The integrated cross section is used to compute both the location and amount of energy deposited.  The stark contrast between the deposition pattern in \texttt{EMstandard} and in the Bichsel model is illustrated in Fig.~\ref{fig:schematic}.  While the energy deposited at each point can vary in \texttt{EMstandard}, the distance between collisions is essentially fixed.  The new straggling function has significant implications for position resolution and is a better model of the data.
 
In CMS, though the Bichsel model has not been used explicitly, their pixel simulation makes use of some smearing factors on Pixelav\cite{swartz2002detailed} which does include the Bichsel model. Hans Bichsel had also done various studies for STAR TPC\cite{Bichsel:2006} and other experiments have also studied the Bichsel model in the past \cite{needham_simulating_2003}. However, there is no general-purpose community tool and there have not been any systematic studies of the impact of track resolution for various thicknesses.

This paper implements the Bichsel model into a standalone Geant4 package called Allpix\cite{benoit:20xx} which is a common tool used for testbeam simulation.  Section~\ref{sec:imple} reviews the model physics and provides a description of the technical implementation.  The simulation framework and comparison metrics between our implementation, \texttt{EMstandard}, and \texttt{PAI} are described in Sec.~\ref{sec:compare} and the numerical results are shown in Sec.~\ref{sec:resul}.  This comparison includes an evaluation of the energy loss, position resolution, and CPU timing.  The paper concludes in Sec.~\ref{sec:concl} with outlook for the future.

	

\section{Bichsel Model Implementation}
\label{sec:imple}
According to the convolution method in ~\cite{Bichsel:1988if}, fluctuations in the energy loss of the high energy particles traversing silicon are mainly due to two sources.  The first source is the number of collisions the particle undergoes inside the material and the second source is the energy loss distribution per collision. The number $n$ of inelastic scatters inside material over length $x$ follows a Poisson distribution:

\begin{align}
\label{eq:NbClo}
	\Pr(n)=\frac{(x/\lambda)^{n}}{n!}e^{-x/\lambda},
\end{align}	

\noindent where $\lambda$ is the mean free path, calculated from the collision cross section $\sigma(E)$ and the number of scattered centers per unit volume $N$ as

\begin{align}
\lambda^{-1} = N\int dE\sigma(E).
\end{align}

\noindent The spectrum for energy loss $\Delta$ after $n$ collisions is calculated by the $n$-fold convolution of single collision spectrum $\sigma(E)$:
\begin{align}
	\label{eq:xsection}
	\sigma(\Delta)^{*n}=\int_{0}^{\Delta}\sigma(E)\sigma^{*(n-1)}(\Delta-E)dE,
\end{align}

\noindent with $\sigma(\Delta)^{*0}=\delta(\Delta)$ (Dirac $\delta$-function) so that  $\sigma(\Delta)^{*1}=\sigma(\Delta)$.  Therefore, the full straggling function is:
\begin{align}
	\label{eq:stra}
	f(\Delta,x)=\sum_{n=0}^{\infty}\frac{(x/\lambda)^{n}e^{-x/\lambda}}{n!}\sigma(\Delta)^{*n}.
\end{align}

\noindent Equation~\ref{eq:stra} does not admit a closed-form analytic solution, but numerical calculations are provided in~\cite{Bichsel:1988if}.  Our implementation follows that of~\cite{qithesis} using a Monte Carlo simulation of the individual scatters inside the sensor. For an incident particle with velocity $\beta\gamma$,  a path length $x$ is sampled according to an exponential distribution whose average is the mean free path.  If moving along the particle trajectory by this amount is still inside the sensor, an energy $E$ is sampled from the probability distribution $\sigma(E)/\int_0^{E_\text{max}(\beta\gamma)}dE'\sigma(E')$.  If this sampled energy is less than the incident particle energy, then the particle is advanced by $x$ and an energy $E$ is recorded.  This process is then repeated until the particle is no longer in the sensor.    Knock-out electrons ($\delta$-rays) are provided by Geant4 (using \texttt{EMstandard}) and occur $\mathcal{O}(1\%$) of the time per 100 $\mu$m of path length in silicon.  To avoid double-counting, the energy loss spectrum in the Bichsel model is cut-off at the Geant4 $\delta$-ray production threshold (chosen to be 117 keV).  Also in the Bichsel model, it is possible that no energy is deposited if the sensor is sufficiently thin.  The probability for depositing any energy in a 1 $\mu$m thick sensor is 97.8\% and the the probability that this energy is above 10~eV is about 86.9\%.  

To save time, a lookup table\footnote{This table-making code came directly from Hans Bichsel.  It has been translated into C++ and is available with the digitizer described in this paper.} stores the cumulative energy distribution over a wide range of $\beta\gamma$ values  and for various particle types including $e,\pi,\mu,P$ and $K$.  In practice, the code interpolates between $\beta\gamma$ points.  In addition to interpolation, two other time-saving measures are implemented: $dE$ bin merging and charge chunking~\cite{qithesis}.  Since the energy is sampled using the cumulative distribution, using large $dE$ bins does not lead to a significant compromise in accuracy.  Charge chunking treats $n$ collisions happening at once.  For thicknesses below $10$ $\mu$m, this feature is not used and for thicknesses above $10$ $\mu$m, 10 collisions are grouped as one. 





\section{Comparison Metrics}
\label{sec:compare}

Numerical simulations using silicon sensors with a variety of thicknesses are studied in order to compare the Bichsel, \texttt{PAI}, and \texttt{EMstandard} charge deposition models.  Two key observables used for this investigation are the energy loss distribution function and the reconstructed position resolution.  The pixel position resolution is critical for all further steps of track reconstruction in a complete detector.

\begin{figure}[h!]
	\includegraphics[width=0.95\textwidth]{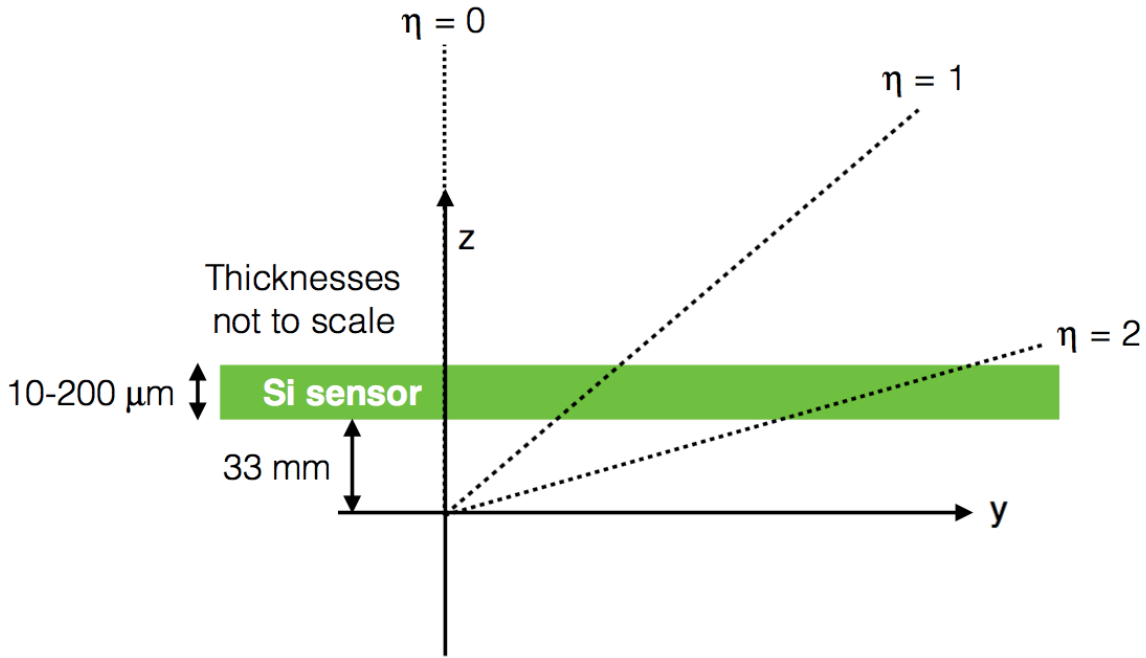}
	\caption{The simulation setup used for the numerical studies. The detector thickness varies between $10$ and $200$ $\mu $m. A $25$ GeV monochromatic muon source is placed at the origin of space. The muon travels in the $+zy$ plane The mean value of the particle pseudorapidity $\eta$ is investigated at discrete values in $\{0,1,2\}$ and $\phi=\arctan(y/x)=0$}
	\label{fig:setup}
\end{figure}

For demonstration, the energy deposition and position resolution is studied for a single layer of a silicon detector with various thickness, illustrated in Fig.~\ref{fig:setup}.  The pitch is fixed at $50\times50$ $\mu \text{m}^2$ and the thickness varies between $10$ and $200$ $\mu $m. A $25$ GeV monochromatic muon source is placed at the origin of space with a rate of one particle/event. The muon travels in the $+zy$ plane at an angle $\theta$ from the $y$ axis.  The one-layer detector is an effectively infinite plane in $xy$ that is $33$ cm in $z$ away from the source. The mean value of the particle pseudorapidity\footnote{By definition, $\eta=-\ln\left(\tan\left(\frac{\theta}{2}\right)\right)$.} $\eta$ is investigated at discrete values in $\{0,1,2\}$ and $\phi=\arctan(y/x)=0$.  

The charge modification described in Sec.~\ref{sec:imple} is implemented in Allpix v1.0\footnote{Technically, the model is implemented during digitization: the \texttt{EMstandard} model is used like normal, but for non $\delta$-rays, the information from Geant4 is replaced with the standalone model.}.  For comparison, Geant4.10.3 is used with either the \texttt{EMstandard} or \texttt{PAI} model.  In a silicon pixel detector, particles may go through one or several pixels, leaving different amount of energy in every pixel. The resulting ionized charge is discretized using a 4-bit Time-over-Threshold (ToT) linear ADC method. The ToT distribution in pixels is then used as input to a fully connected artificial neural networks (NN) to estimate the entrance point of the incident particle~\cite{Hocker:2007ht,Aad:2014yva}. The position residual $P_\text{estimated} - P_\text{truth}$ measures the accuracy of the estimation, and is used to compare the three models: Bichsel, \texttt{PAI} and \texttt{EMstandard}.

\section{Results}
\label{sec:resul}

\begin{figure}[h!]
\centering
	\includegraphics[width=0.4\textwidth]{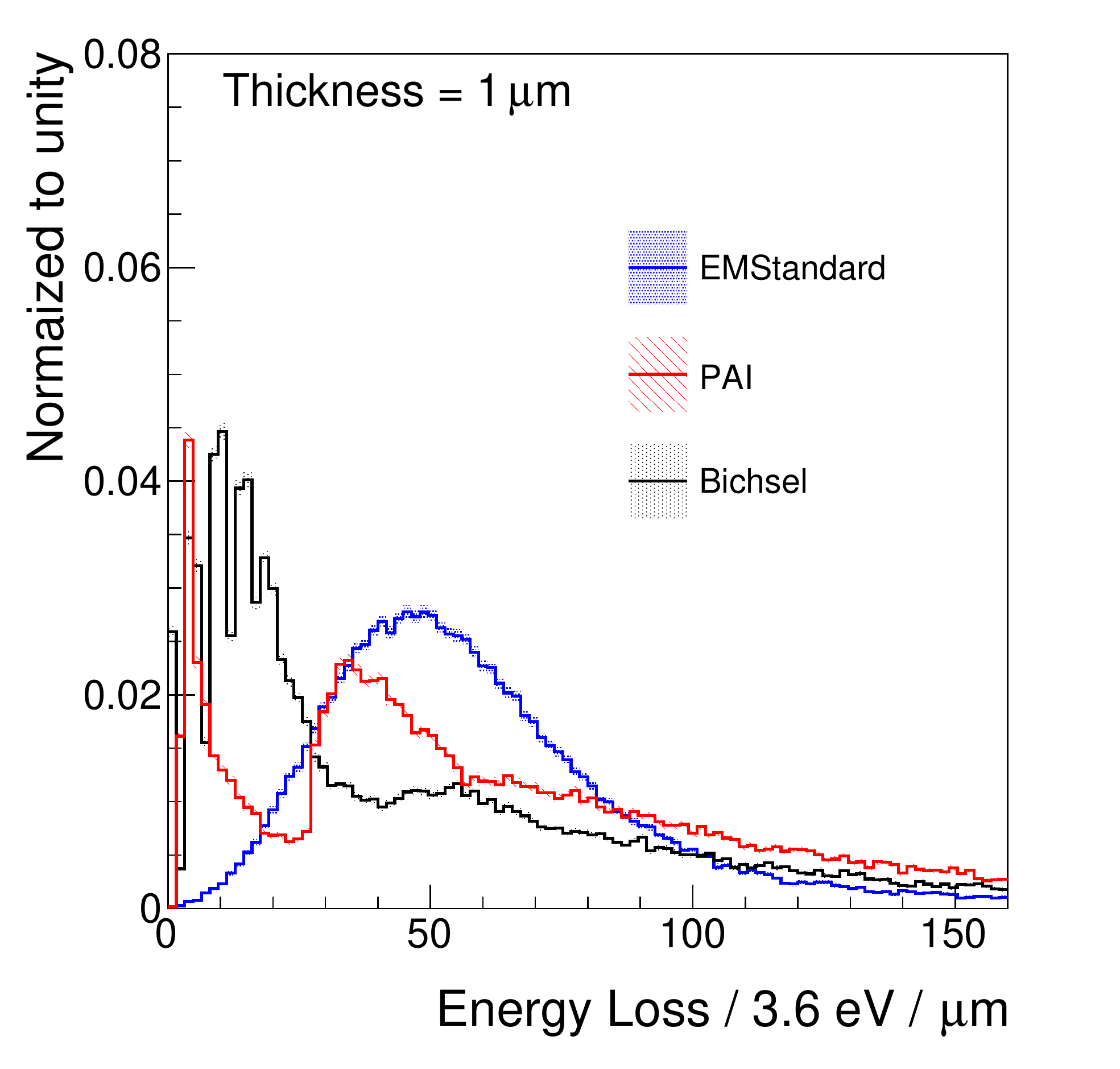}
	\includegraphics[width=0.4\textwidth]{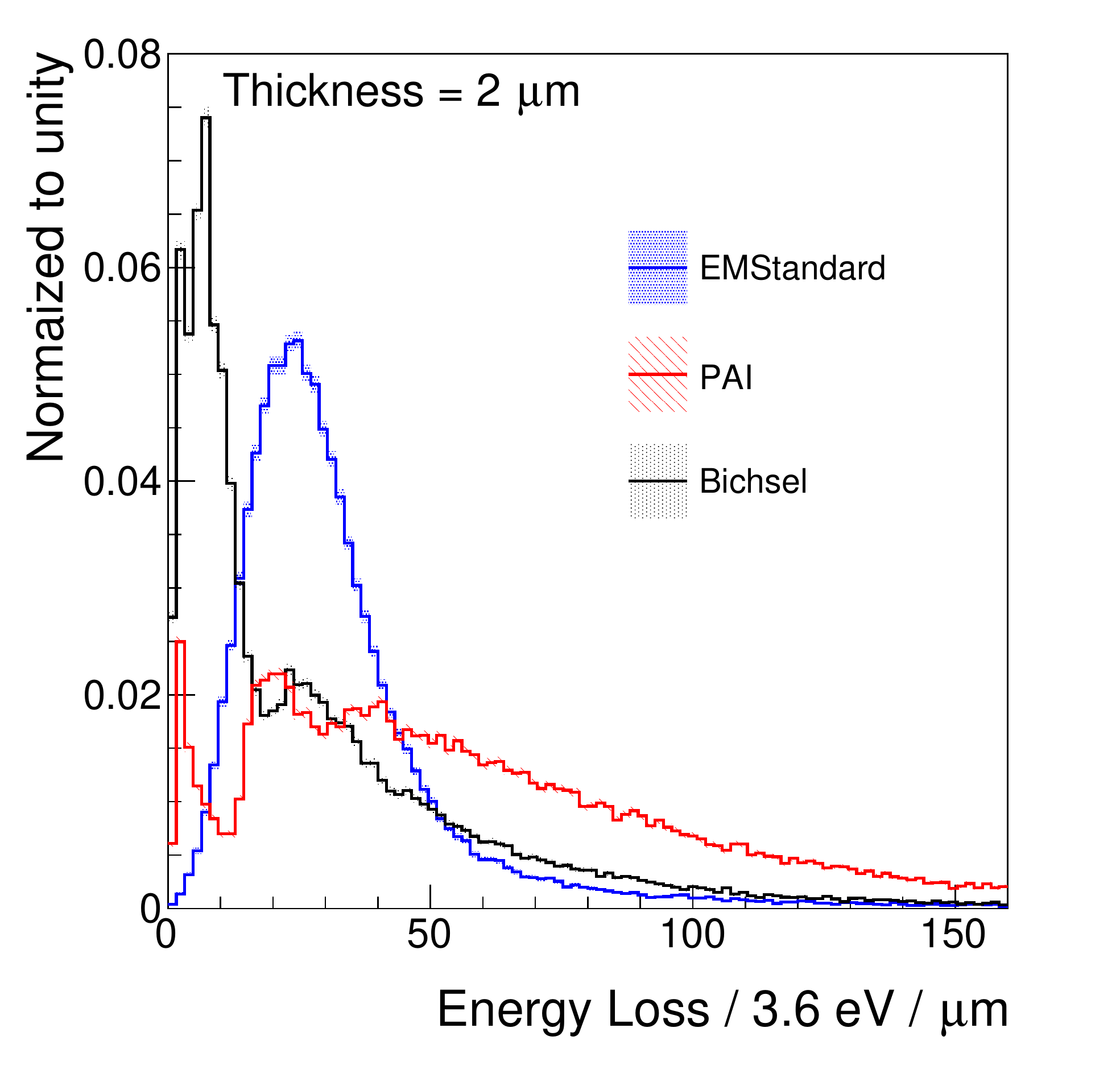}\\
	\includegraphics[width=0.4\textwidth]{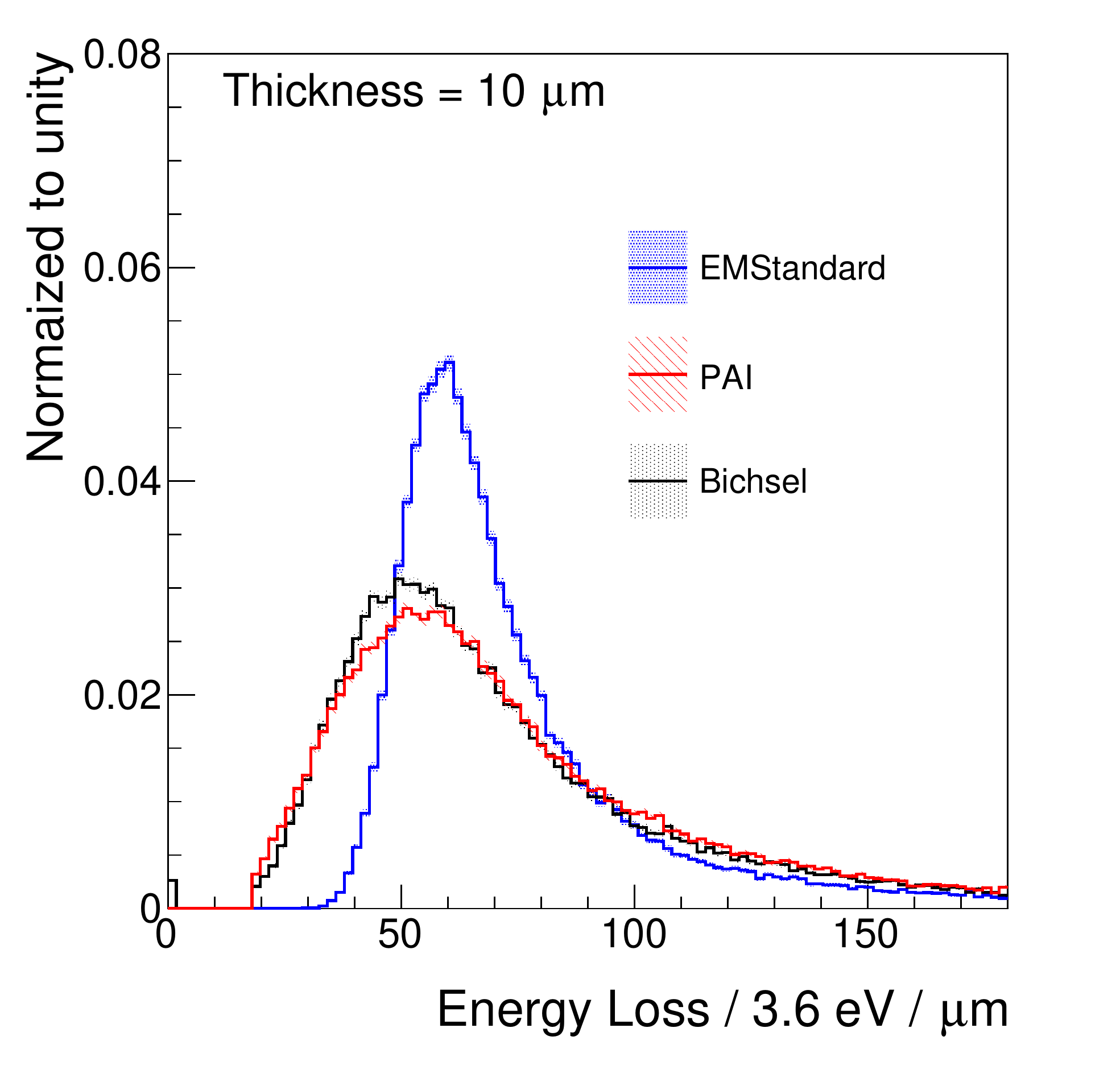}
	\includegraphics[width=0.4\textwidth]{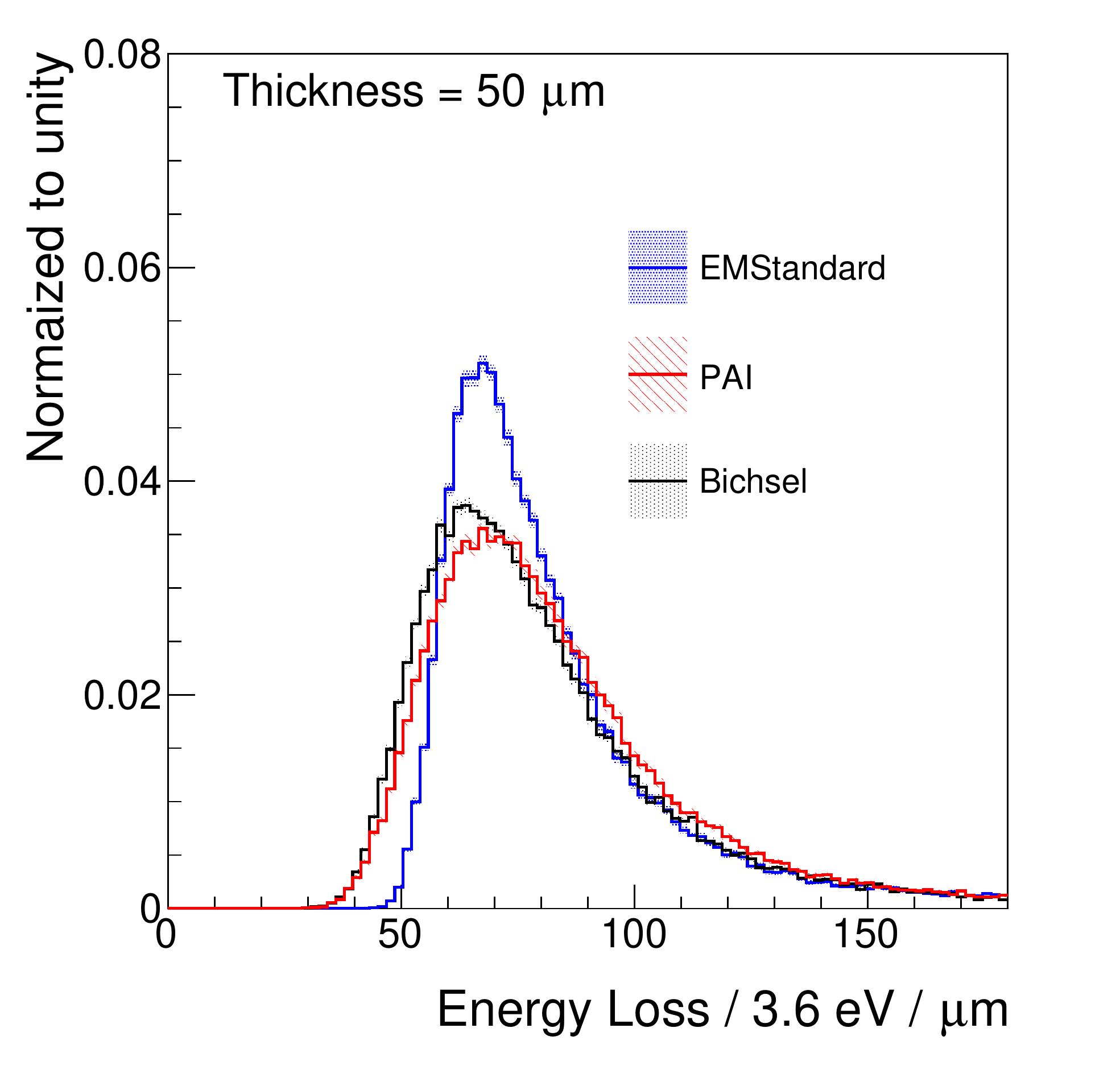}\\
	\includegraphics[width=0.4\textwidth]{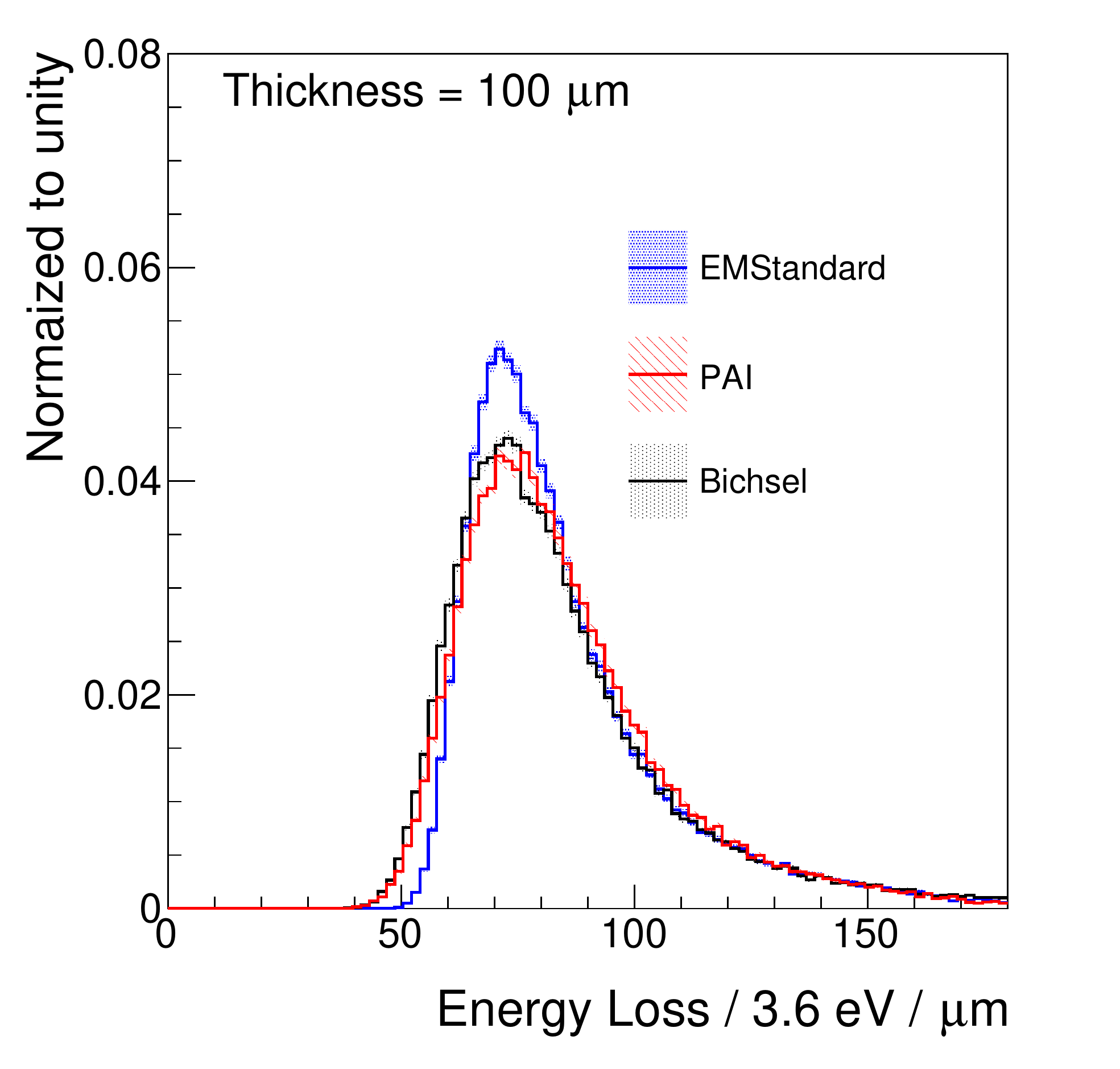}
	\includegraphics[width=0.4\textwidth]{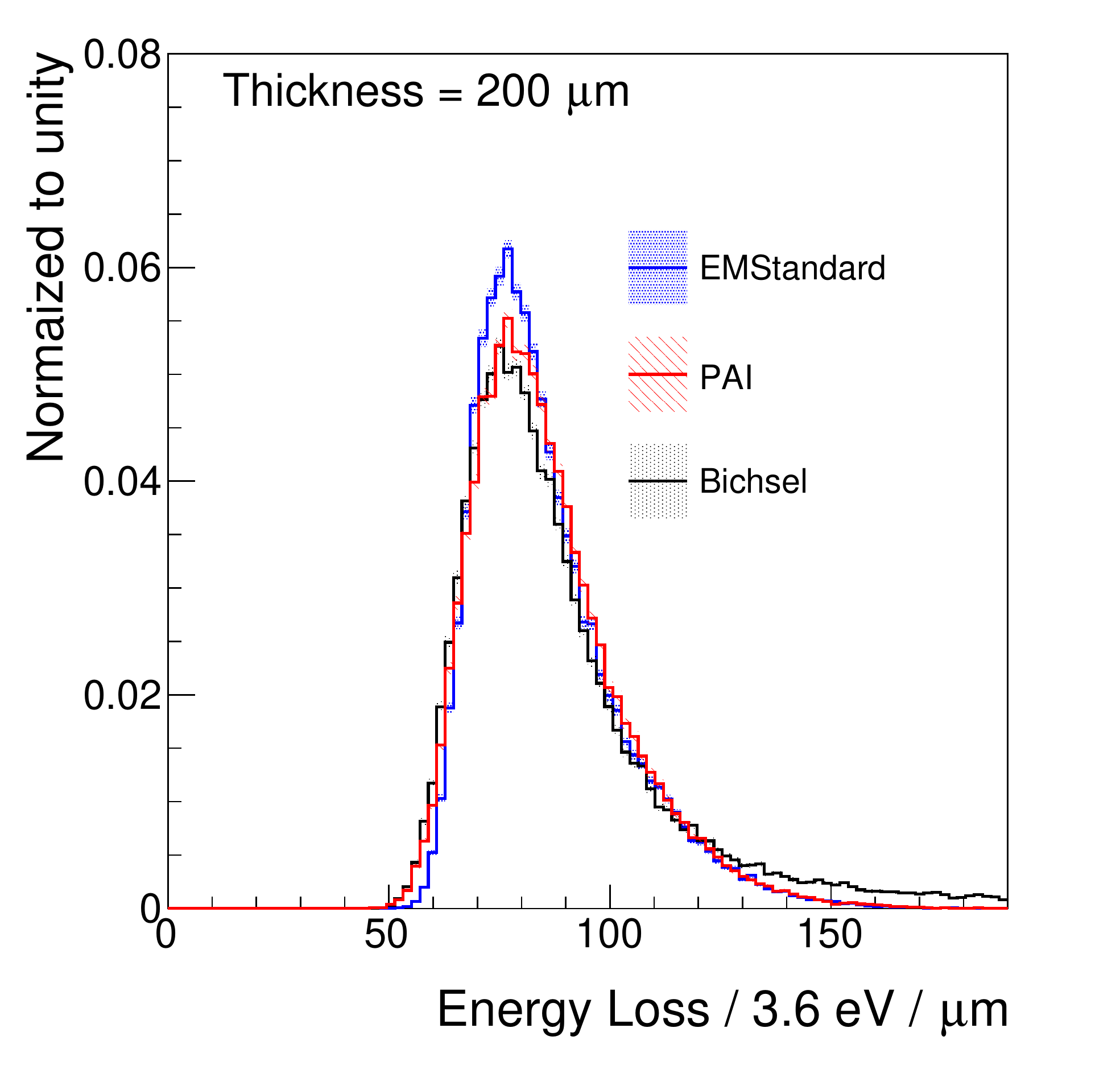}
	\caption{The average energy loss per micron for the Bichsel, \texttt{PAI} and \texttt{EMstandard} straggling models for sensors that are $1$ $\mu$m (top left), $2$ $\mu$m (top right), $10$ $\mu$m (middle left), and $50$ $\mu$m (middle right) thick, $100$ $\mu$m (bottom left), and $200$ $\mu$m (bottom right) thick.} 
	\label{fig:energy}
\end{figure}


The average energy loss per micron with a variety of detector thicknesses is shown in Figure \ref{fig:energy}. The energy loss distribution for PAI and Bichsel models are similar because both of these two models include shell corrections, which considers the actual electronic structure of the atoms\cite{Bichsel:1988if}. The difference with \texttt{EMstandard} is large, especially when the detector thickness is less than 10 $\mu$m. The shell-corrected models have a similar peak position to  \texttt{EMstandard}, though the spread in energy values is larger.  The agreement between the three models is good when the thickness reaches $200$ $\mu$m, though there are still some small differences in the width.  Below $10$ $\mu$m, there are large differences between all three models.  The discrete energy levels from shell effects are very pronounced in the Bichsel model.  The multi-bump structure in both the Bichsel and \texttt{PAI} models is from the transition between one, two, or multiple collisions inside the silicon.  The width enhancement in the \texttt{EMstandard} for very thin sensors is clearly visible when comparing $1$ and $2$ $\mu$m.

\begin{figure*}[h!]
	\centering
	\includegraphics[width=0.55\textwidth]{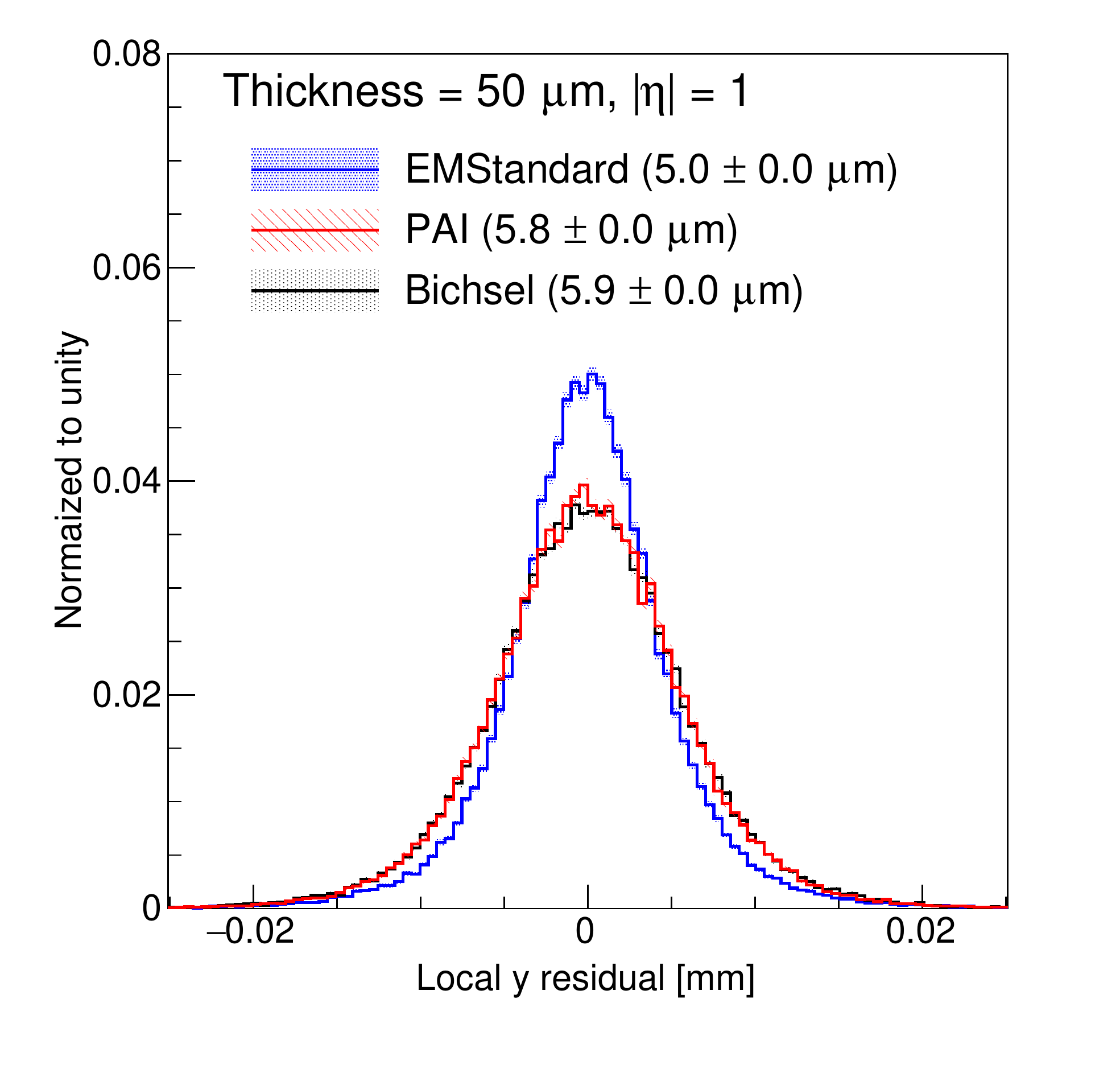}
	\caption{The difference between the predicted and true position of a 25 GeV muon passing through a 50 $\mu$m thick silicon detector at $\eta=1$ for Bichsel, \texttt{PAI}, and \texttt{EMstandard} models.}
	\label{fig:threemodel}
\end{figure*}

\begin{figure*}[h!]
	\includegraphics[width=0.5\textwidth]{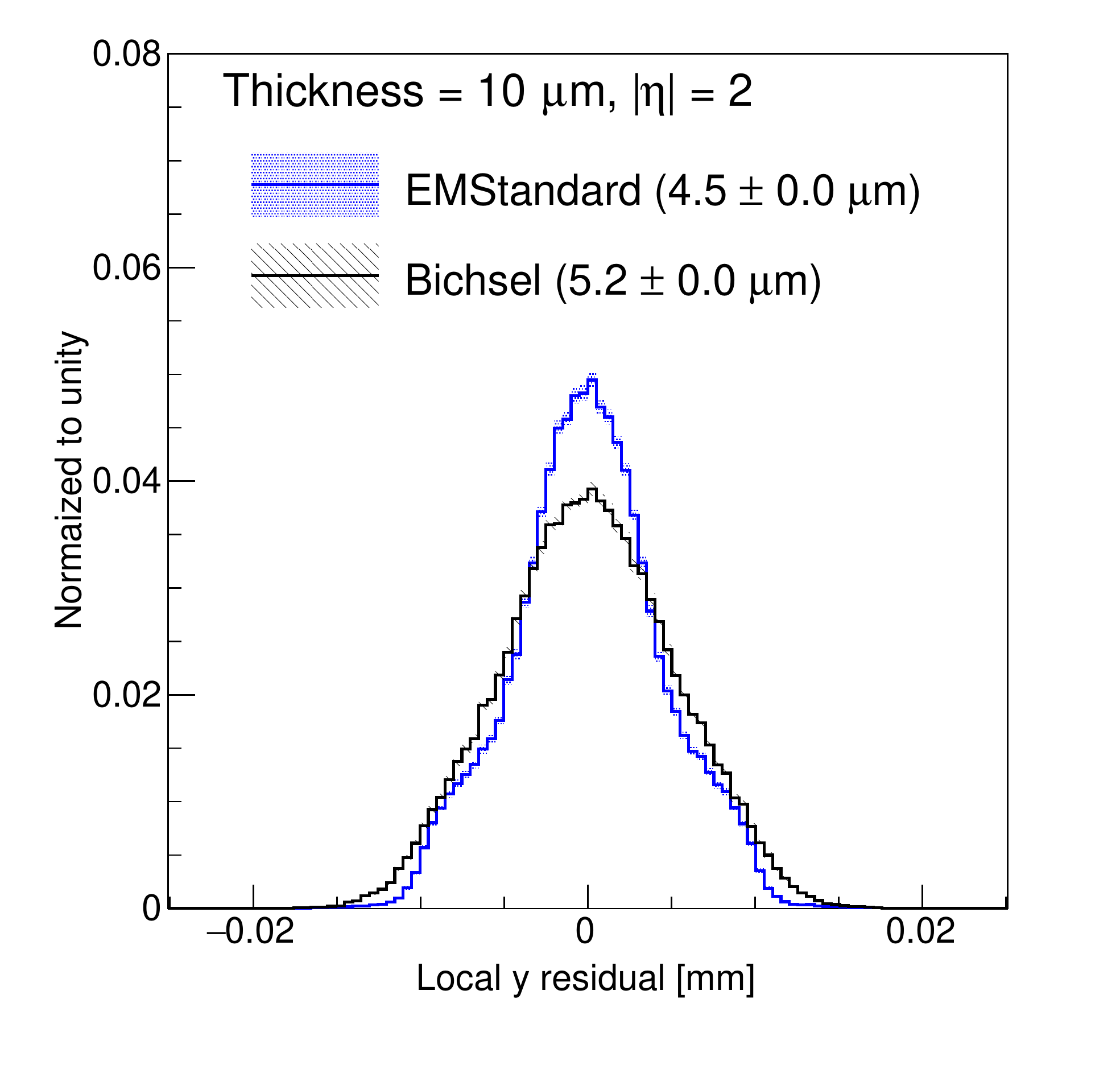}
	\includegraphics[width=0.5\textwidth]{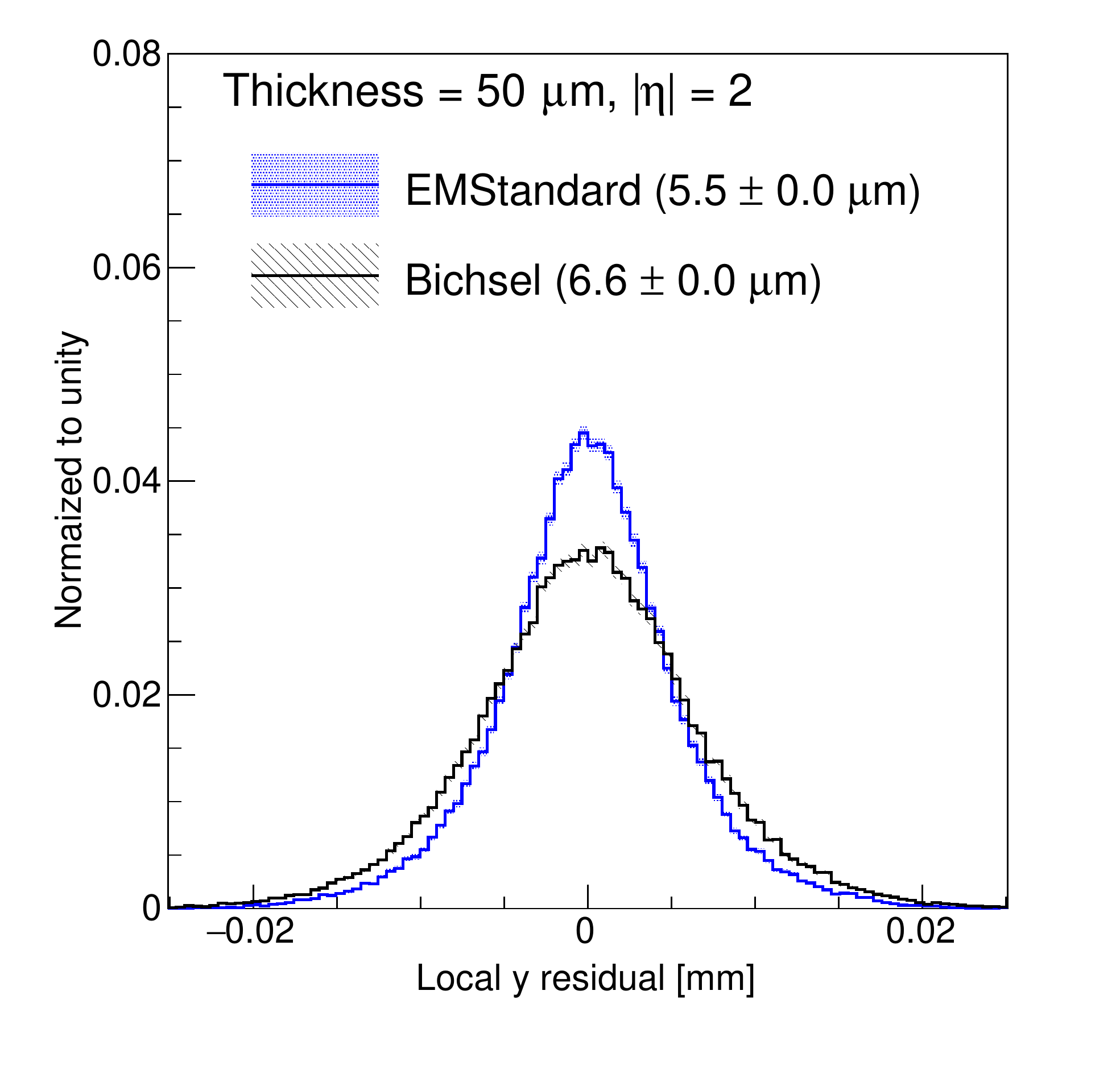}
	\includegraphics[width=0.5\textwidth]{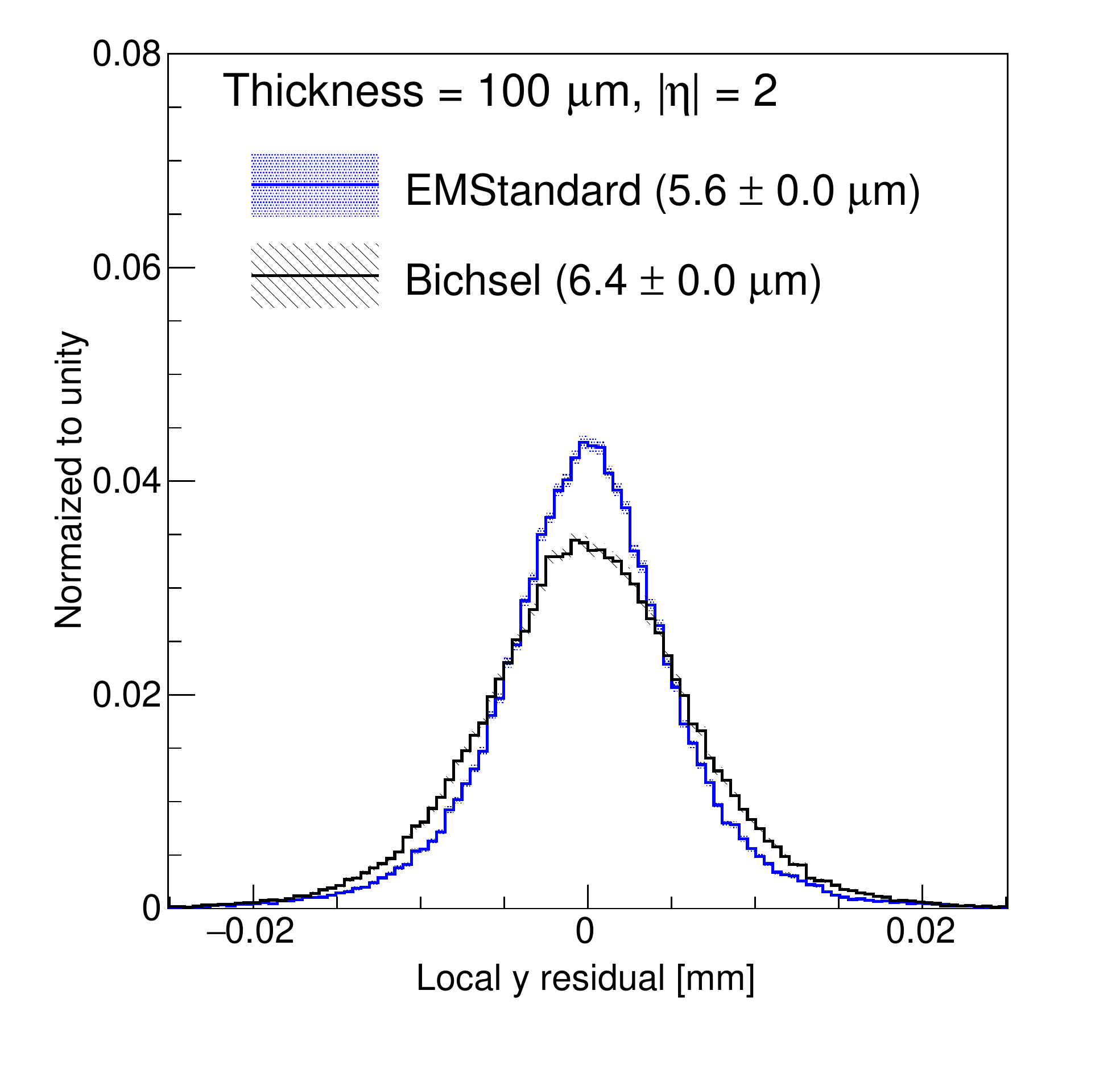}
	\includegraphics[width=0.5\textwidth]{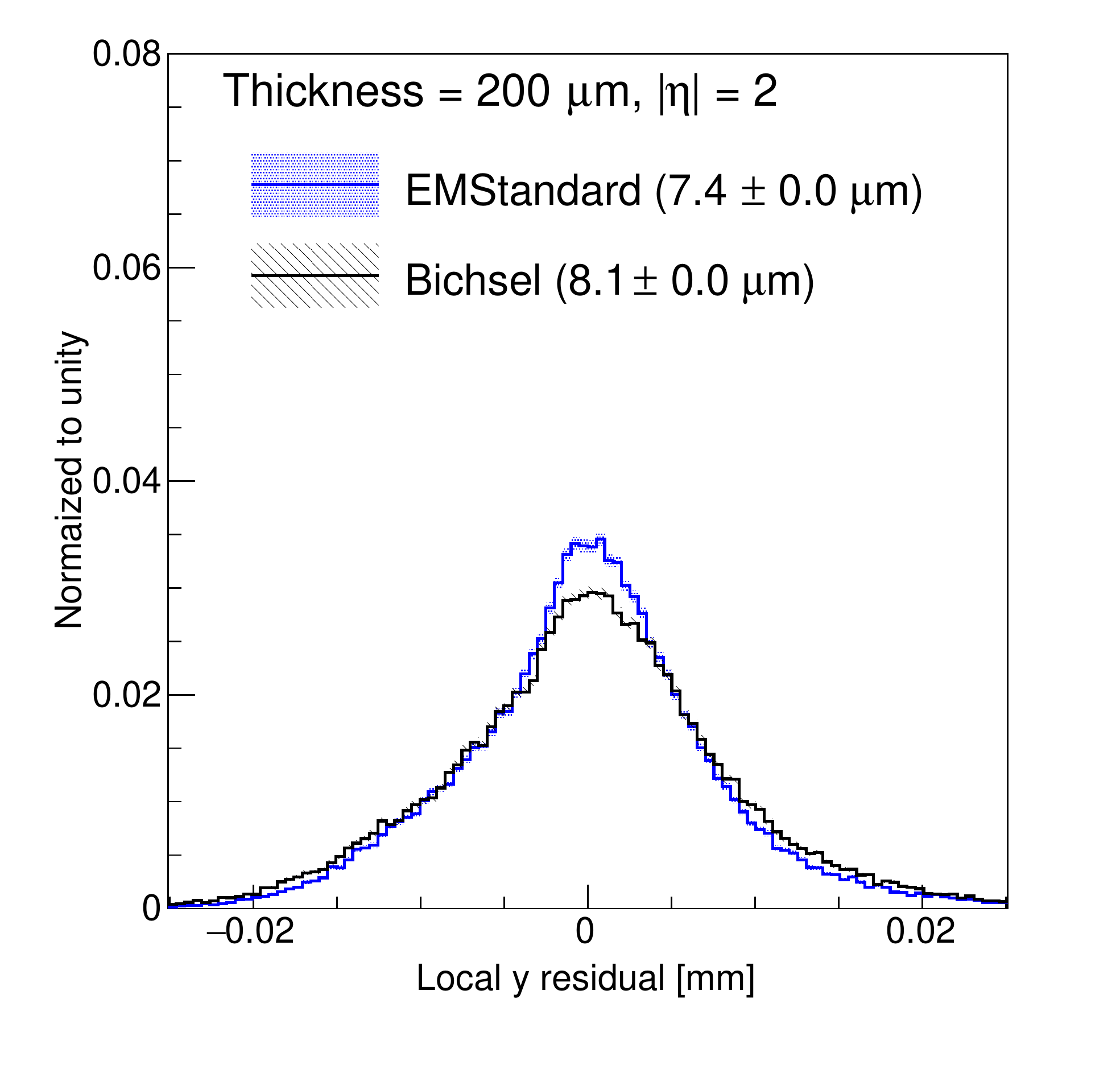}
	\caption{Position residual distribution $\eta=2$. When thickness is 10 $\mu$m, the shape of \texttt{EMstandard} shows sign of two different pixel clusters, while for Bichsel, the wide distribution of the residual makes this unclear. The discrepancy between Bichsel and \texttt{EMstandard} is visible even for 200$\mu$m-thick-detectors.}
	\label{fig:twomodel}
\end{figure*}

Variations in the energy loss distributions between straggling functions lead to differences in the reconstructed particle hit clusters.  For example, Fig.~\ref{fig:threemodel} shows the position residual of the Bichsel, \texttt{PAI} and \texttt{EMstandard} when $\eta=1$ and the thickness is 50 $\mu$m. The residual distributions of the Bichsel and  \texttt{PAI} models agree well with each other, but once again have a larger standard deviation compared to \texttt{EMstandard}.  This has important implications for improving the modeling of pixel clusters as well as higher level objects including jet tagging~\cite{qithesis,ATL-PHYS-PUB-2017-013}.  Comparisons for additional sensor thicknesses at $\eta=2$ are shown in Fig.~\ref{fig:twomodel} for the Bichsel and \texttt{EMstandard} models (Bichsel and \texttt{PAI} are nearly the same). When the detector thickness is 10 $\mu$m, the shape of \texttt{EMstandard} shows sign of two different pixel clusters, while for Bichsel, the wide distribution of the residual makes this unclear. The discrepancy between Bichsel and \texttt{EMstandard} is visible even for 200$\mu$m-thick-detectors. 



\begin{figure*}[h!]
	\centering
	\includegraphics[width=0.5\textwidth]{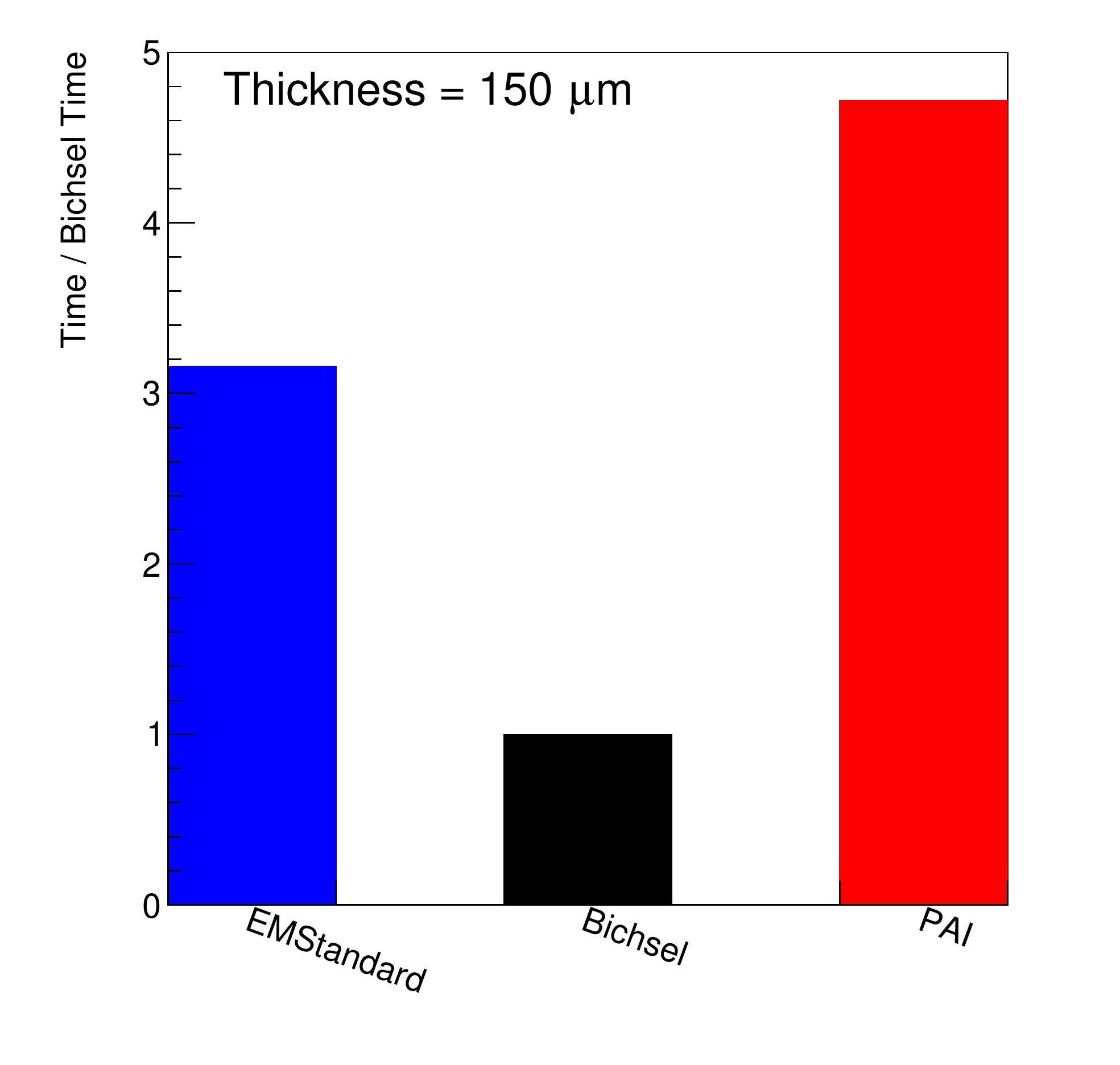}
	\caption{The total time for charge deposition over many events normalized to unity for the Bichsel model and a 150 $\mu$m thick sensor.}
	\label{fig:time}
\end{figure*}

Even though it is nearly identical to the Bichsel model, the \texttt{PAI} is not used in practice due to its long simulation time.  Figure~\ref{fig:time} compares the CPU time when simulating the same number of events under the three different straggling models.  The \texttt{PAI} model is about 50\% slower than \texttt{EMstandard} while the implementation of the Bichsel model is about three times faster than \texttt{EMstandard}.  Additional savings may be possible, at the cost of accuracy with coarser bins in the $\sigma(E)$ lookup.

\section{Conclusions}
\label{sec:concl}

The energy loss of high energy particles passing through thin detectors is not well-described by the classical Landau straggling function. This paper implements the Bichsel model~\cite{Bichsel:1988if} into a standalone simulation framework Allpix~\cite{benoit:20xx} that can be used directly for testbeam studies and readily ported to other frameworks\footnote{The code will be made available at \url{https://github.com/LBNLPixelProjects}.}.  The Bichsel straggling function is compared with the most widely used energy deposition model in Geant4~\cite{AGOSTINELLI2003250}, \texttt{EMstandard}, as well as a more sophisticated model that includes similar physical effects as the Bichsel model (\texttt{PAI}).  The energy deposition fluctuations significantly differ between \texttt{EMstandard} and the other two models, which are themselves similar.  Differences in straggling have a significant impact on reconstruction, as demonstrated by the resolution of the reconstruction position residual.  Even though the Bichsel and \texttt{PAI} models have similar energy deposition patterns, the implementation of the Bichsel model presented here is three times faster.  

As silicon sensors become thinner, it will be increasingly important to properly model the non-Landau nature of the straggling function.  The algorithm presented here will empower simulation and modeling studies for thin sensors at testbeams and beyond.



\section{Acknowledgments}


We would like to thank Chris Damerell for spotting a minor inconsistency in the caption of Fig. 3.  This work was supported by the U.S.~Department of Energy, Office of Science under contract DE-AC02-05CH11231 and by the China Scholarship Council. 

\clearpage

\bibliographystyle{elsarticle-num}
\bibliography{myrefs.bib}{}

\end{document}